\documentclass[aps,twocolumn,showpacs,floatfix,amssymb,preprintnumbers,prl]{revtex4-1}

\usepackage[dvips]{graphicx}
\usepackage{dcolumn}
\usepackage{bm}
\usepackage[dvips]{epsfig}
\usepackage{amssymb}
\usepackage{latexsym}
\usepackage{hyperref}
\newcommand \be {\begin{equation}}
\newcommand \ee {\end{equation}}
\newcommand \bea {\begin{eqnarray}}
\newcommand \eea {\end{eqnarray}}

\begin{document}

\title{Softening induced instability of a stretched cohesive granular layer}

\author{Hector Alarc\'on$^{\dagger}$, Osvanny Ramos, Lo\"ic Vanel$^\ddagger$, Franck Vittoz, Francisco Melo$^{\dagger}$ and Jean-Christophe G\'eminard.}
\affiliation{Universit\'e de Lyon, Laboratoire de Physique, Ecole Normale Sup\'erieure de
Lyon, CNRS, UMR 5672, 46 All\'ee d'Italie, 69007  Lyon, France.}
\affiliation{$^{\dagger}$ Departamento de F\'{\i}sica, Universidad de Santiago de Chile,
Av. Ecuador 3493, Casilla 307, Correo 2, Santiago, Chile.}
\affiliation{$^\ddagger$ Universit\'e de Lyon, Universit\'e Lyon 1, Laboratoire de Physique Mati\`ere Condens\'ee and Nanostructures, CNRS, UMR 5586, 69622 Villeurbanne, France}

\begin{abstract}
We report on a cellular pattern which spontaneously forms at the surface of a thin layer of a cohesive
granular material submitted to in-plane stretching. We present a simple model in which the mechanism responsible of the instability is the ``strain softening'' exhibited by humid granular materials above a typical strain.
Our analysis indicates that such type of instability should be observed in any system presenting a negative stress sensitivity to strain perturbations.
\end{abstract}

\maketitle

\textsl{Introduction -- }
Adding even minute amounts of liquid can change dramatically the mechanical properties of
sand.
During the building of sand castles,
one observes a transformation from a fluid-like
to a sticky and deformable material
with increasing water content.
Indeed, at very low water content, the formation of partially-developed capillary-bridges
leads to a fast increase of tensile strength whereas, for large enough fluid content,
tensile strength is nearly constant \cite{Fournier_2004}.
Cluster formation was identified as the main mechanism responsible of such a behavior \cite{Scheel_2008}.

One unexplored but important feature of the mechanical response of cohesive sand, arising in a wide range of fluid content, is that the tensile strength decreases when the imposed strain is increased.
Such a ``strain softening'' is due both to a decrease of the associated adhesion force when a single bridge is elongated \cite{Crassous_97} and to a decrease in the overall number of bridges which collapse when excessively stretched \cite{Willett_2000}.
We mention that the softening behavior is only observed above a critical strain which vanishes
for sufficient stiff grains \cite{groger03}. Indeed, the critical strain is associated with the initial
compression of the grains induced by the suction force due to the capillary bridges.

In the present Letter,
we report that, in a stretched layer of cohesive grains, ``strain softening'' induces a mechanical instability in which the strain field is modulated in space.
Our analysis indicates that such type of instability should be observed in any system presenting a negative stress sensitivity to strain perturbations.

\begin{figure}[!h]
\begin{center}
\includegraphics[width=\columnwidth]{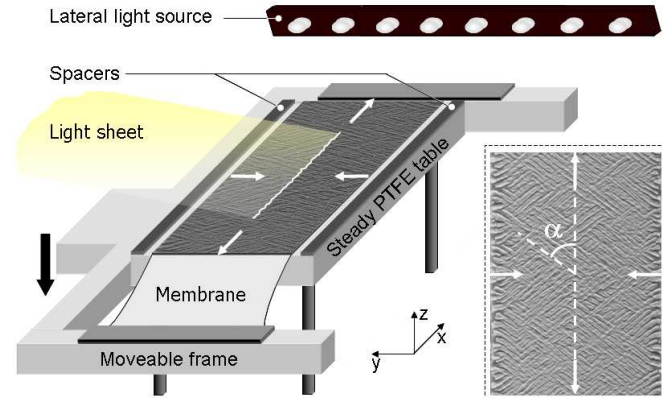}
\end{center}
\caption{\small{Sketch of the experimental setup --} Inset: Top view of the whole pattern and definition of the angle $\alpha$.}
\label{setup}
\end{figure}

\textsl{Experimental setup and protocol -- }
The experiment consists in imposing an in-plane deformation at the base of a thin layer of a cohesive granular material.
To do so, the grains are initially spread onto an elastic membrane to which the deformation is imposed (Fig.~\ref{setup}). The membrane, a thin Latex band (thickness 0.5 mm, width 5 cm and total length 20 cm) is maintained at its two ends by a moveable U-shaped frame and leans, in its central part, on a steady rectangular table (length 10 cm, width 6 cm).  By displacing the frame downwards, the band, which remains in the same horizontal plane above the table, extends along its length and narrows in the perpendicular direction (Note that, due to the contact with the table, the membrane does not wrinkle).
The top plate of the table is made of polytetrafluoroethylene (PTFE), which limits the friction and insures a reproducible deformation for a given frame displacement.

The sample is prepared first by pouring grains
onto the membrane. The surface of the material is then leveled by means of a cylindrical rod guided
by lateral spacers, which achieves a well-defined thickness $h$ (from 1 to 5~mm, to within 0.1~mm).
The granular material consists of spherical glass-beads (USF Matrasur, sodosilicate glass).
We shall report results obtained for various samples in a large range of bead diameters $d$
(0-45, 45-90,
100-125
~$\mu$m).
In order to tune the cohesion, the experimental device is placed in a chamber
in which the atmosphere is equilibrated with a saturated salt solution.
The cohesion is, in addition, accounted for by measuring the angle of avalanche $\theta_a$ \cite{bocquet98,Nowaki_2005}
in the same experimental conditions (A granular layer, same material and thickness, is prepared onto a rough surface which can be tilted).

The free surface of the sample is imaged from above by means of a digital camera (Konica-Minolta, A200).
Two linear light-sources (home-made arrays of LEDs, Fig.~\ref{setup}) placed at the two ends of the elastic band,
about 20~cm away from the sample, a few centimeters above the table plane, provide a good contrast when the
upper surface of the material is deformed.
In addition, in order to assess quantitatively the vertical displacement of the free surface,
we cast with an angle of 30~deg, a light sheet onto the sample.
In this configuration, the horizontal displacement of the bright line is proportional to the local
vertical displacement of the free surface.

\begin{figure}[!h]
\begin{center}
\includegraphics[width=.8\columnwidth]{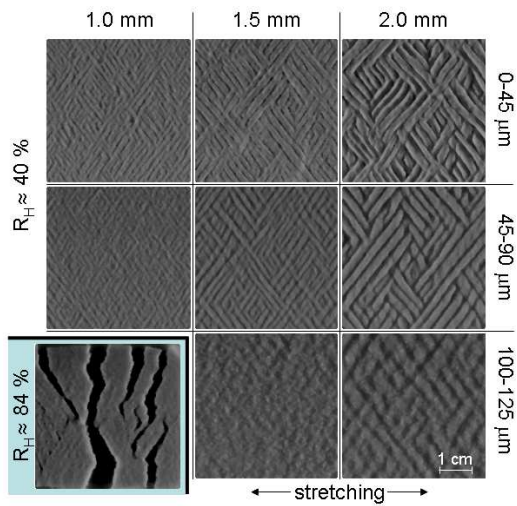}
\end{center}
\caption{\small{{ Surface structure for distinct layer thickness and particles size at constant stretching --}
Lower left panel : a highly cohesive layer slides on the membrane
(Labels: Top, $h$; Right, $d$; $\theta \simeq 0.2$) }}
\label{illustration}
\end{figure}

\textsl{Results -- }
When the membrane is stretched by moving the frame downwards,
one observes, provided that the grains are small enough and/or the relative humidity, $R_H$,
large
enough, the growth of a complex pattern at the free surface of the granular layer (Fig.~\ref{illustration}).

Domains, made of stripes having a rather well-defined width and making an angle
$\pm \alpha$ with the stretching direction ($x$-axis,  Fig.~\ref{setup}), nucleate and grow.
We note that the phenomenon consists of the continuous growth of a free-surface undulation, which,
depending on the experimental conditions, can eventually lead to the fracture of the granular layer when the stretching is further increased.
Note that the undulation first appears at the edges of the latex band
where it makes a $90$~deg angle with the stretching direction (Fig.~\ref{setup}, inset), which
indicates,
first, that the uniaxial stretching can be responsible, alone, for the instability, and, second, that the compression along the $y$-axis plays a role in the angle $\alpha$ measured in the central part of the sample.
We can also deduce from the observation of domains, disconnected from the edges, exhibiting a given orientation ($\pm \alpha$)
that the overall pattern does not result from the growth of the instability occurring at the lateral boundaries.

\begin{figure}[!h]
\begin{center}
\includegraphics[width=.8\columnwidth]{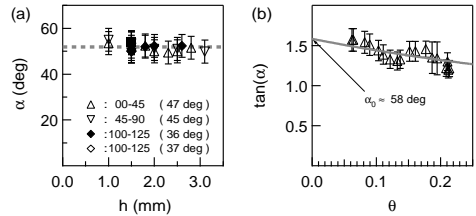}
\end{center}
\caption{\small{{(a) Angle $\alpha$ vs. thickness $h$ for $\theta \simeq 0.21$ --} The angle $\alpha$ neither depends on the
thickness $h$ of the granular layer nor on the grain size $d$ and cohesion [Symbol : d ($\theta_a$)].
{(b) Angle $\alpha$ vs. elongation $\theta$ --} The stretching induces a rotation of the stripes such that $\tan(\alpha)={(2-\theta)}/{[2(1+\theta)]}\,\tan(\alpha_0)$. From the extrapolation of the data  (grey line), one assesses the angle $\alpha_0 \simeq 58$~deg ($d = $0-45$~\mu$m, $h = 2$~mm and $\theta_a = 47$~deg).}}
\label{angle}
\end{figure}

Let us first discuss the angle $\alpha$.
We denote $\theta \equiv u_{xx}$, the strain in the $x$ direction ($\vec{u}$ denotes the displacement field)
and report $\alpha$ measured for a given $\theta$ (Fig.~\ref{angle}a).
We observe that, to within the experimental accuracy ($\pm 3$~deg on one sample),
$\alpha$ neither depends on $h$, $d$ or cohesion (accounted for by $\theta_a$)
and we get $\alpha = (51.5 \pm 0.5)$~deg for $\theta\simeq0.21$.
In contrast, $\alpha$ depends on $\theta$. Indeed, the latex being almost incompressible, $u_{yy}\simeq-\theta/2$
and the stripes are expected to rotate according to $\tan(\alpha)=\frac{2-\theta}{2(1+\theta)}\,\tan(\alpha_0)$.
Reporting $\tan(\alpha)$ as a function of $\theta$ (Fig.~\ref{angle}b), we get $\alpha_0 \simeq 58$~deg,
the value of $\alpha$ in the limit $\theta \to 0$.
One can account for the experimental angle $\alpha_0$ by considering the Mohr criterium \cite{nedderman_1992}.
On the one hand, the stretching tends to pull the grains apart,
so that the normal stress along the $x$-axis, $\sigma_{xx} \lesssim \sigma_s$,
where $\sigma_s$ denotes the tensile stress. On the other hand, the compression
pushes the grains one against another and one can guess that the associated stress
involves the solid contacts between the grains and, consequently, that $|\sigma_{yy}|\gg |\sigma_{xx}|$.
Thus, one can consider that the granular material is subjected to a pure compressive stress along the $y$-axis.
With this assumption, the Mohr criterium leads to $\alpha_0 = \frac{\pi}{4}+\frac{\Phi}{2}$
where $\tan{(\Phi)}=\mu$, the static friction coefficient \cite{nedderman_1992}.
The experimental value of $\alpha_0$ corresponds to $\Phi \simeq 26$~deg
associated with $\mu \simeq 0.5$, a reasonable value of the static friction coefficient
for a packing of glass spheres \cite{friction}.

\begin{figure}[!t]
\begin{center}
\includegraphics[width=.8\columnwidth]{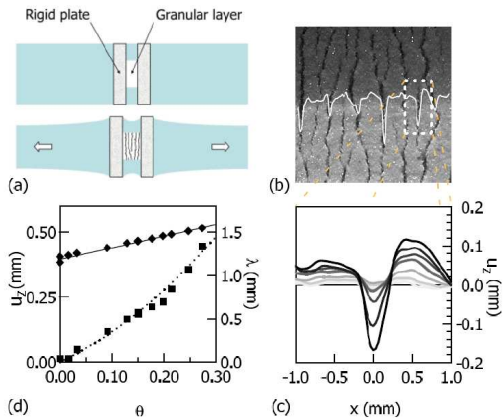}
\end{center}
\caption{\small{{(a) Sketch of the experimental configuration for uniaxial strain.}
{(b) Photograph of the fracture pattern -- }
Fractures are perpendicular to the strain axis.
The white line is the vertical displacement $u_z(x,h)$ of the surface.
{(c) Vertical displacement of the free surface $u_z(x,h)$ -- }
The depth at the fracture increases whereas the free surface is observed to rise, and thus the granular layer to dilate, between the fractures when $\theta$ is increased (A darker line indicates a larger $\theta$).
{(d) Wavelength $\lambda$ (diamonds) and amplitude (squares) of the
vertical undulation $u_z(x,h)$ vs. strain $\theta$ -- } The amplitude increases continuously with $\theta$ whereas, due to the simple advection of the pattern,
 $\lambda$ obeys $\lambda_0 (1+\theta)$ ($d = 100$-125$~\mu$m, $h = 2$~mm and $\theta_a = 37$~deg).}}
\label{uniaxial}
\end{figure}

We measured the typical width of the stripes and amplitude of the vertical deformation of the layer.
We observed that similar results are obtained when a pure uniaxial deformation of the membrane is imposed.
Thus, seeking for simplicity, we shall report measurements associated with the deformation
of the layer in the simplest geometry: we limit the experiment to 1-cm-wide membrane stripe
(length 5 cm), limited by two thin metal plates glued onto the membrane (Fig.~\ref{uniaxial}a).

When the membrane is stretched, the central region, away from the edges,
is subjected to a pure uniaxial strain and one then observes the formation
of stripes perpendicular to the elongation axis (Fig.~\ref{uniaxial}b).
The deformation of the bright line casted onto the surface (Fig.~\ref{setup})
makes it possible to obtain the vertical profile of free surface along a line and
thus to quantify the growth of the pattern (Fig.~\ref{uniaxial}c).
We observe that, obviously, the fracture is marked by a sharp decrease of
the surface altitude but, interestingly, that the height of the free surface
in the domains between the fractures increases and, thus, that the in-plane
stretching leads to a thickening of the layer.

\begin{figure}[!h]
\begin{center}
\includegraphics[width=.8\columnwidth]{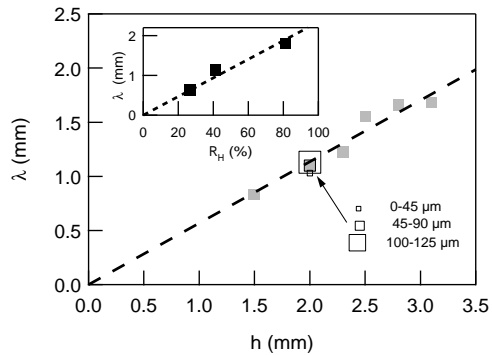}
\end{center}
\caption{{Wavelength $\lambda$ vs. thickness $h$ -- }
Full squares: $\lambda$  is proportional to $h$.
For $d = 45$-90$~\mu$m, $\lambda \simeq 0.6\,h$;
Open squares: $\lambda$ does not
significantly depend on $d$ ($\theta \simeq 0.3$ and $R_H = 35$\%).
 For clarity, error bars ($20$\% in $\lambda$, 0.1 mm in $h$) are not shown. --
Inset: $\lambda$ increases with $R_H$
($\theta \simeq 0.3$, $d = 100$-125$~\mu$m and $h = 2$~mm).}
\label{table1}
\end{figure}

At this point, it is interesting to consider the dependency of the typical wavelength, $\lambda$,
of the fracture pattern on the experimental parameters that are $h$, $d$ and $R_H$. Note first
that $\lambda$ is not strictly selected and that we observe a large scatter of the stripes
width. In spite of the scatter, we observe that $\lambda \propto h$ (Fig.~\ref{table1})
as long as $h$ does not exceeds about 3~mm (If $h$ is too large,
the bands between the fracture are likely to split in two, leading to a smaller average $\lambda$).
For a given $R_H$ and $h$, $\lambda$ is almost independent of $d$ (Fig.~\ref{table1}) whereas one observes a significant increase of $\lambda$ when cohesion is increased by increasing $R_H$ (Inset: Fig.~\ref{table1}).

\textsl{Theoretical analysis -- }
The instability requires cohesion.
Due to the nature of the interaction between the grains,
the adhesion force decreases when the material is stretched and,
thus, grains are pulled apart \cite{Willett_2000,groger03}.

Qualitatively, when the membrane is elongated, an homogeneous stretching
of the material is imposed in the base plane.
However, due to the ``strain softening'', in response to the overall stretch,
the system tends spontaneously to modulate the deformation: regions of large
deformation are associated with a smaller modulus and regions of large modulus
are associated with a smaller deformation, which results in an overall decrease of the
energetic cost. In turn, the modulation induces a shear deformation which is associated
to an energetic cost. Thus, the wavelength is governed by the balance of the gain
associated with the modulation of the horizontal strain and of the loss associated
with the induced shear. We point out here, that the shear (the relative motion of the top and bottom surfaces)
is consistent with the dilation of the layer in the vertical direction (Fig.~\ref{uniaxial}c).

In order to account, at least qualitatively, for the experimental observations
let us first note that the normal stress along the $x$-axis, $\sigma_{xx}$,
decreases linearly with the uniaxial strain $u_{xx}$, according to
$\sigma_{xx} = \sigma_s ( 1 - {u_{xx}}/{\theta_m})$ when the material
is stretched ($u_{xx} >0$) \cite{groger03}. The relation is no longer valid for
$u_{xx} > \theta_m$, when the elongation is large enough for the bridges to collapse
and, thus, the material to break apart.
Thus, $\theta_m$ is of the order of the typical size of the bridge $\delta$
divided by the grains diameter $d$ whereas $\sigma_s$ denotes the tensile stress
previous to deformation.
For the sake of simplicity, the contribution of the shear shall be accounted for
by a simple shear modulus $G$ which value shall be discussed later.
In this framework, the shear stress $\sigma_{xz} = G u_{xz}$ and, accordingly, the energy per unit volume
\begin{equation}
E = \sigma_s \Bigl(u_{xx} - \frac{u_{xx}^2}{2 \theta_m}\Bigr)+\frac{1}{2} G u_{xz}^2.
\end{equation}
We thus assume that the dilation in the vertical direction $u_{zz}$
does not contribute to any additional energetic cost: The grains are assumed to remain
in contact along the vertical and no significant stretching of the capillary bridges occurs
in this direction.

We consider now a sinusoidal perturbation of the displacement such that $u_x = \theta x + f(z) \sin{(k x)}$
in the horizontal plane. In order to obtain the associated displacement in the vertical direction $u_z$,
seeking for simplicity, we will further assume that the deformation of the material does not
induce any dilation so that $u_{xx}+u_{zz}=0$.
Writing the relation imposed by the mechanical equilibrium in the horizontal plane
$\frac{\partial \sigma_{xx}}{\partial x}+\frac{\partial \sigma_{xz}}{\partial z}=0$ \cite{Landau_1959}
and the conditions that $u_x$ and $u_z$ do not depend on $x$ at the substrate plane ($z = 0$), we get
$u_x = \theta x + a \omega \sin{(\omega z)}\sin{(k x)}$ and
$u_z = -\theta z - a k [1-\cos{(\omega z)}]\cos{(k x)}$
where $a$ is an amplitude and $\omega^2 = k^2 (1+\xi)$ with $\xi\equiv {2\sigma_s}/{(G \theta_m)}$.
Writing that the shear stress $\sigma_{xz}$ vanishes at the free surface and, thus, that $u_{xz} = 0$
for $z=h$, we get $1+\xi\cos{(\omega h)} = 0$. The wavelength $\lambda \equiv 2\pi/k$ is thus found to be
proportional to $h$, independent of $\theta$, according to
\begin{equation}
\lambda = 2\pi\frac{\sqrt{1+\xi}}{\arccos{(-1/\xi)}}h
\label{longueur}
\end{equation}
provided that $\xi \ge 1$. One can check that in the accessible range of $\xi$, the energy $E$ is a decreasing
function of the amplitude $a$, whatever the strain $\theta$. Thus, the layer is always unstable provided
that the decrease in the tensile stress is large enough compared to the shear cost,
i.e. $\frac{\sigma_s}{\theta_m} \ge \frac{G}{2}$.
The growth of the instability is limited by the condition that $u_{xx}(x,h) \ge 0$ for all $x$ at the free surface
(the strain in the plane $z=0$ does not lead to any compression in the plane $z=h$),
so that $\theta- a \omega k \sin{(\omega h)}=0$. Thus, the amplitude of the vertical displacement,
$k\,a$, is predicted to be proportional to $\theta\,h$.

\textsl{Discussion -- }
Interestingly, the theoretical analysis, which only involves a decrease of the tensile stress
associated with the stretching of the material and an energetic cost associated with the induced
shear, predicts that a stretched layer is always unstable.
In agreement with the experimental observations, the instability
does not exhibit any finite threshold, the amplitude of the modulation increases
linearly with $\theta$ for small strain (Fig.~\ref{uniaxial}d) and
the wavelength $\lambda$ is proportional to the layer thickness h (Fig.~\ref{table1}).
In addition, the amplitude is predicted to be proportional to $\theta h$,
which explains why the pattern is more easily observed when the layer is thick
and the stretching is large.

For larger $\theta$, the instability develops further and the deformation is no more sinusoidal.
The stretching concentrates in small regions in which we estimate that
the relative displacement of two grains is of the order of $\theta \lambda$.
When $\theta \lambda \simeq \delta$, the typical size of a bridge, the material is locally torn off.
Then, instead of an undulation of the free surface, one rather observes fractures.
Thus, taking into account that $\lambda \propto h$ and that $\delta \propto d$, one predicts that thick layers or small grains, which are associated with very small critical strain $\sim d/\lambda$, always lead to fractures, in agreement with the experimental observations.

Finally, we note that the effect of the humidity content on the wavelength is accounted for by the dependence of $\lambda$ on the ratio $\xi \equiv -(d\sigma/du_{xx})/G$. For instance, in the limit of large $u$ (small bridges),
$\lambda \simeq 4 \sqrt{2\sigma_s/G\theta_m}h$. The increase of $\lambda$ with $R_H$ would impose, in the framework of the crude model,
that $G$ increases slower than the ratio $\sigma_s/\theta_m = -(d\sigma/du_{xx})$ (Note that $\sigma_s$ and $\theta_m$ both increase with $R_H$). However, the peculiar choice of a linear elastic response of the material to shear is certainly a matter of debate and we do no discuss this point further.
We nevertheless point out that the most important features are not model dependent:
the instability has in practice no threshold, the wavelength scales like the thickness and the amplitude
of the undulation increases linearly with both the stretching and the layer thickness, .

\textsl{Conclusions -- }
We reported for the first time the destabilization process of a material deposited onto
an elastic membrane which exhibits a negative stress-strain sensitivity, $d\sigma/du_{xx}<0$, or equivalently ``strain-softening'',
which generally leads to the fracturing of the material.
However, as we demonstrated, the process differs significantly from the usual fracturing processes
\cite{Corson_2009, Ikeda_2008} as the instability develops gradually with increasing external strain.
Interestingly, the instability mechanism reported here applies to any system having a negative sensitivity to stretching and, thus, is not expected to be specific to the granular matter.

The authors acknowledge the financial support from the contracts FONDAP 11980002, MECESUP USA-108, ANR-05-JCJC-0121-01 and ECOS-CONICYT C07E04.

\end{document}